\documentclass[prd,twocolumn,floatfix,nofootinbib,superscriptaddress]{revtex4}
\usepackage[applemac]{inputenc}
\usepackage{amsmath}
\usepackage{amssymb}
\usepackage{amsthm}
\usepackage{bm}
\usepackage{epsfig}
\usepackage{graphicx}
\usepackage{longtable}
\usepackage{color}
\usepackage{epstopdf}
\usepackage{nicefrac}
\usepackage{tabularx}
\usepackage{slashed}
\usepackage{multirow}
\usepackage{dcolumn}  
\newcolumntype{L}[1]{>{\raggedright\arraybackslash}p{#1}}
\newcolumntype{C}[1]{>{\centering\arraybackslash}p{#1}}
\newcolumntype{R}[1]{>{\raggedleft\arraybackslash}p{#1}}

\usepackage[colorlinks=true, pdfstartview=FitV, linkcolor=purple, citecolor= purple,urlcolor=blue]{hyperref}
\definecolor{darkgreen}{rgb}{0,0.5,0}
\definecolor{purple}{rgb}{0.5,0,0.5}
\definecolor{nblue}{rgb}{0.0,0.0,0.50}
\definecolor{scarlet}{rgb}{1.0,0.2,0}

\def\nn{\nonumber} 
\def\Tr{\operatorname{Tr}}

\def\k{\bm{k}}

\usepackage{soul}

\graphicspath{{Figures/}}

\begin{document}

\title{Parton distribution functions and transverse momentum dependence of heavy  mesons}

\author{Fernando E. Serna}
\affiliation{Departamento de F\'{\i}sica, Universidad de Sucre, Carrera 28 No. 5-267, Barrio Puerta Roja, Sincelejo, Colombia}

\author{Bruno El-Bennich}
\affiliation{Departamento de F\'{\i}sica, Universidade Federal de S\~ao Paulo, Rua S\~ao Nicolau 210, Diadema, 09913-030 S\~ao Paulo, Brazil}
\affiliation{Instituto de F\'{\i}sica Te\'orica, Universidade Estadual Paulista, Rua Dr. Bento Teobaldo Ferraz, 271 -- Bloco II, 01140-070 S\~ao Paulo, SP, Brazil}

\author{Gast\~ao Krein}
\affiliation{Instituto de F\'{\i}sica Te\'orica, Universidade Estadual Paulista, Rua Dr. Bento Teobaldo Ferraz, 271 -- Bloco II, 01140-070 S\~ao Paulo, SP, Brazil}


\begin{abstract}
The leading Fock state light-front wave functions of heavy quarkonia and $D$ and $B$ mesons are obtained from the projections of their Bethe-Salpeter wave functions on the light front. We compute therefrom their leading-twist time-reversal even transverse momentum distributions and parton distribution functions. Mirroring the behavior of parton distribution amplitudes, the support of both distributions is increasingly narrower and shifted towards larger $x$ as a function of the meson mass. The dependence of $x$ and $\k_\perp^2$ of the transverse distributions does not factorize into separate functions, and their fall-off with $\k_\perp^2$ is much slower than that of light mesons.
\end{abstract}

\date{\today}

\maketitle


\section{Introduction}

The definition of Light-Front Wave Functions (LFWFs) plays a fundamental role in the light-front quantization approach to Quantum Chromodynamics (QCD)~\cite{Brodsky:1997de}, where 
they are essential objects that encode the nonperturbative internal dynamics of hadronic bound states. They provide a frame-independent representation of hadrons and a unique perspective 
on the \emph{longitudinal and transverse} dynamics  of their constituents. One can derive probability amplitudes from them, which describe the probability of 
finding partons, quarks and gluons with a given longitudinal fraction of the hadron's total momentum as well as their transverse momentum dependence. 
Two of these distributions, the parton distribution function (PDF) and transverse momentum distribution (TMD), define hadronic operators that relate the partons to measurable hadronic 
observables~\cite{Jaffe:1996zw,Burkardt:2002uc, Pasquini:2014ppa,Metz:2016swz}, as they are experimentally accessible and provide valuable insights into the internal structure of hadrons.

From the perspective of continuum QCD and its application to hadron phenomenology, a LFWF provides a starting point to compute generalized parton distributions, form factors, 
structure functions, and gravitational form factors. On the other hand,  since the light front is not directly accessible in nonperturbative approaches formulated in Euclidean space, 
the calculation of a hadron's LFWF poses a challenge. This is certainly the case in lattice-regularized QCD and with functional methods in continuum QCD, and has prompted 
alternative approaches to solving light-front quantized field theories directly~\cite{Pauli:1985ps,Vary:2009gt}, for instance with a large momentum expansion of the equal-time 
Euclidean correlation functions in instant quantization~\cite{Ji:2021znw} or with the projection of the Bethe-Salpeter wave function on the light front~\cite{Mezrag:2016hnp,Shi:2018zqd}. 
A recent method described in \cite{Eichmann:2021vnj} introduces a contour-deformation technique combined with analytic continuation methods, offering a promising way to project the 
Bethe-Salpeter wave function directly onto the light front. These approaches collectively contribute to addressing the challenges of computing LFWFs in a Euclidean framework.

The  approaches based on light-front projection have recently been pursued to extract the LFWF from an Euclidian wave function computed within the combined framework 
of the Dyson-Schwinger equation (DSE) and Bethe-Salpeter equation (BSE)~\cite{Bashir:2012fs} and to compute the PDF and TMD of light pseudoscalar and vector mesons 
~\cite{Shi:2018zqd,Shi:2020pqe,Kou:2023ady,Shi:2023oll}. Likewise, the LFWFs of $D$ and $B$ mesons were extracted from Mellin moments using an algebraic ansatz for the 
Bethe-Salpeter wave function in  Ref.~\cite{Almeida-Zamora:2023bqb}. Herein, we fill a gap and extend these studies to the LFWF of $D$, $D_s$, $B$, $B_s$ and $B_c$ mesons 
with a DSE-BSE calculation  that preserves the axialvector Ward identity~\cite{Rojas:2014aka,Mojica:2017tvh,Serna:2020txe,Serna:2021xnr,Serna:2022yfp,daSilveira:2022pte}. 
With the wave functions in hand  we then compute the meson's TMD and PDF.  


\section{Light front wave functions\label{LFWFsec}}

The minimal $\bar{q} q$ Fock-state configuration of a pseudoscalar meson in light-front QCD is given by~\cite{Burkardt:2002uc},
\begin{equation}
   \left | M(P)\right\rangle=\left| M(P) \right \rangle_{l_{z}=0} + \left | M(P)\right\rangle_{\left|l_{z} \right  |=1} .
\end{equation}
The two states are described by two independent LFWFs, namely $\psi_{0}\left(x, \k^2_\perp\right)$ for the angular projection $l_{z}=0$ and $\psi_{1}\left(x, \k^2_\perp\right)$ 
for $\left | l_{z}\right|=1$, where $ \k_\perp$ is the transverse momentum and $x=k^+ / P^+$ is the light-front momentum fraction of the quark. In a frame where the meson's transverse 
momentum vanishes, $\bm{P}_\perp = 0$, the transverse momentum of the antiquark is $\bar \k_\perp = - \k_\perp$ and its light-front momentum fraction is $\bar x =1-x$.  
The light-front components of a Euclidean four-vector $k_\mu = (k_1,k_2,k_3,k_4)$  are defined by $k^\pm = ik_4\pm k_3$ and in terms of a light-like vector, $n^2=0$, we have
 $k^{+}=n\,\cdotp k$. Finally, the transverse momentum of the quark is defined by $\k_\perp=(k_1,k_2)$ .

Starting from the matrix element definitions of the LFWFs, one can show that the pion's minimal Fock-state LFWFs is obtained from its Bethe-Salpeter wave function
$\chi_M(k, P)$, with the two projections~\cite{Mezrag:2016hnp},
\begin{align}
 \Psi_{l_z}(x,\k^2_\perp )  =  & \sqrt{3}i\int\frac{dk^+dk^{-}}{2\pi}\, \delta\! \left (x P^+ - k^+_\eta \right )  \nn   \\[0.2true cm]
     & \times  {\Tr}_D \left [ \mathcal{O}^{+5}_{l_z}  \chi_M(k,P)  \right ] \, ,
 \label{LFWFs-01}
\end{align}
where $\mathcal{O}^{+5}_{l_z=0}=\gamma^+\gamma_5$ and ${\cal O}^{+5}_{|l_z|=1}=-i\sigma_{+i}k^i_\perp\gamma_5/\k^2_\perp$ and  \linebreak the  trace is over Dirac indices. 
The Bethe-Salpeter wave function is defined by 
\begin{equation} 
\chi_M(k,P) = \int d^4 z\ e^{-ik \cdot z}\, \langle 0|\mathcal{T} f(z)\,\bar g (0)| M(P)\rangle \,, 
\end{equation} 
and can be expressed as,
\begin{equation}
  \chi_M(k,P) = S_f(k_\eta)\,\Gamma^{fg}_{M}(k,P)\,S_g(k_{\bar\eta} ) \, ,
 \label{BSAwavefunc}
\end{equation}
with the relative and total momenta $k$ and $P$, respectively. The momenta of the dressed quark propagators $S_f(k_{\eta})$ of given flavor $f$, $k_\eta =k+\eta P$ and 
$k_{\bar\eta}=k-\bar\eta P$, define momentum-fraction parameters:  $\eta+\bar\eta =1$. We recall that observables cannot  depend on them due to translational invariance. The Bethe-Salpeter amplitude (BSA) of a pseudoscalar meson made of a quark $f$ and antiquark $g$ is generally defined as~\cite{Llewellyn-Smith:1969bcu},
\begin{align}
\label{PS-BSA}
 \Gamma^{fg}_M(k;P) = &\, \gamma_5 \Big  [ iE^{fg}_M(k,P) + \gamma\cdot PF^{fg}_M(k,P)    \nn   \\[0.2 true cm]
    + & \, \gamma\cdot k\,G^{fg}_M(k,P) + \sigma_{\mu\nu}k_\mu P_\nu H^{fg}_M(k,P)\Big ] . 
\end{align}

The wave functions $\chi_M(k,P)$ of pseudoscalar mesons, in particular of heavy-light bound states, are discussed in Refs.~\cite{Serna:2020txe,daSilveira:2022pte} and we refer to 
details therein. They are obtained in the DSE and BSE framework with a self-consistent solution of the quark gap equation~\cite{Serna:2018dwk}, which yields the dressed quark 
propagator for a given flavor, and of the homogeneous BSE, which gives the BSA of the meson. Our numerical solutions, however, are in Euclidean space where do not have direct 
access to objects defined on the light front.  We use the following strategy to derive the LFWF from the Euclidean-space wave functions: we compute $\k^2_\perp$-dependent 
moments of the LFWF  defined by,
\begin{eqnarray}
\label{defxm}
    \langle x^m\rangle_{l_z} (\k^2_\perp ) = \int^1_0dx\,x^m\,\Psi_{l_z}(x,\k^2_\perp ) \ ,
\end{eqnarray}
from which one can reconstruct the wave function on the domain $x\in [0, 1]$ and $\k^2_\perp \in [0,{\Lambda^2_\perp} ]$. Using Eq.~\eqref{LFWFs-01} and integrating over the Dirac function,
one finds, 
\begin{align}
\label{xm-LFWFs-01}
  \langle x^m\rangle_{l_z} (\k^2_\perp ) = & \   \frac{\sqrt{3}i}{|P^+|}  \int  \frac{dk^+dk^{-}}{2\pi} \left(\frac{k^+}{P^+}\right)^{\! m}  \nn     \\[0.3true cm]
    &\times \,  \Tr_D \left [ \mathcal{O}^{+5}_{l_z}\chi_M(k,P) \right ]\, .
\end{align}


\begin{table*}[t!]
\centering
\begin{tabular}{ |C{3cm} ||  C{1.8cm}| C{1.8cm} | C{1.7cm} | C{1.7cm}| C{1.7cm}  |C{1.7cm}| C{1.7cm}  C{1.7cm}  }
\hline \hline
Meson/Parameters &$l_z$ &${\cal N}_M$ & $\Lambda_1$ & $\Lambda_2$ & $\alpha$  &  $\beta$   \\ [0.7mm]
\hline
\multirow{2}{*}{$\pi (u\bar d)$}&0&$13.53\pm2.05$&$1.00\pm0.01$&$1.00\pm0.01$&$0.51 \pm 0.08$ &$0.51 \pm 0.08$ \\
 &$|1|$&$47.08 \pm 8.07$& $1.26 \pm 0.01$ &$1.26 \pm 0.01$&$0.89 \pm 0.10$&$0.89 \pm 0.10$  \\\hline  
\multirow{2}{*}{$K(s\bar u)$} &0 &$24.60 \pm 0.75$ & $1.20 \pm 0.01$ & $1.03 \pm 0.01$ &$1.19\pm0.07$&$0.60 + 0.07$ \\&$|1|$&$85.77 \pm 13.09$&$1.20 \pm 0.02$&$1.18 + 0.01$& $1.37 \pm 0.13$&$1.14 + 0.08$ \\\hline 
\multirow{2}{*}{$D_u(c \bar u)$} &0&$5.82 \pm 1.03$ &$2.13+ 0.06$&$1.06 \pm 0.03$&$-0.29 \pm 0.18$ &$-1.84 \pm 0.12$\\&$|1|$&$16.89 \pm 3.46$ &$1.89 \pm 0.08$&$1.20 \pm 0.03$&$1.10 \pm 0.20$&$-2.47 \pm 0.16$\\\hline 
\multirow{2}{*}{$D_s(c \bar s) $}& 0 & $3.01 \pm 1.00$ &$2.08\pm0.11$&$1.38 \pm 0.06$&$-1.41 \pm 0.35$&$-1.85 \pm0.21$\\ &$|1|$&$5.90 \pm 0.73$&$1.85 \pm 0.18$&$1.69 \pm 0.07$&$-0.33 \pm 0.13$&$-2.36 \pm 0.11$\\\hline 
\multirow{2}{*}{ $\eta_c(c \bar c) $}  &0 &  $0.17 \pm 0.06$  &$2.19 \pm 0.04$&$2.19 \pm 0.04$&$-5.53 \pm 0.40$&0.00\\ &$|1|$&$6.75 \pm 2.22$&$2.11 \pm 0.041$&$2.11 \pm 0.041$&$-1.14 \pm 0.41$&0.00\\\hline 
\multirow{2}{*}{ $B_u(b \bar u) $}   &0&  $10.52 \pm 2.26$&$5.53 \pm 0.14$&$0.99 \pm 0.07$&$2.22 \pm 0.20$&$-2.50 \pm 0.19$\\&$|1|$&$4.98 \pm 0.50$ &$4.79\pm 0.18$&$1.40 \pm 0.05$&$4.58 \pm0.06$&$-4.72 \pm 0.13$\\\hline 
\multirow{2}{*}{ $B_s(b \bar s) $} &0&$13.21\pm 3.45$&$4.86\pm 0.33$&$1.56\pm 0.15$&$1.93 \pm 0.30$&$-2.15 \pm 0.23$
\\&$|1|$&$9.91 \pm 1.18$&$4.95 \pm 0.45$&$1.68 \pm 0.12$&$4.13 + 0.32$&$-4.25 + 0.13$\\\hline 
\multirow{2}{*}{ $B_c(b \bar c) $} &0& $0.61 \pm 0.23$
&$5.17 \pm 0.30$&$2.21 \pm 0.22$&$-3.30 \pm 0.36$&$-3.71 \pm 0.32$\\&$|1|$&
$3.52 \pm 0.90$&$4.91 \pm 0.28$&$2.32 + 0.18$&$-0.69 \pm 0.26$&$-2.61 \pm 0.26$\\\hline 
 \multirow{2}{*}{ $\eta_b(b \bar b) $}  &0& $0.02 \pm 0.01$&$5.47 \pm 0.11$&$5.47 \pm 0.11$&$-8.05 \pm0.28$& $0.00$ \\&$|1|$&$7.97 \pm 2.41$&$5.28 \pm0.12$&$5.28 \pm0.12$&$0.56 + 0.46$& 0.00\\
\hline \hline
\end{tabular}
\caption{\label{tab:Psi-par} Parameters obtained from the LFWF reconstruction via Eq.~\eqref{LFWFs-rec} for the angular momenta $l_z=0$ and $l_z=|1|$.}     
\end{table*}  


To facilitate the calculation of the moments defined by Eq.~\eqref{xm-LFWFs-01},  we use an accurate parametrization of numerical solutions to the quark DSE and BSAs in 
rainbow-ladder truncation. We implement a complex conjugate pole parametrization (\texttt{ccp}) for the quark propagator defined by~\cite{Bhagwat:2002tx,El-Bennich:2016qmb}:
\begin{equation}
\label{ccp}
 S_f  (q )  =  \sum^N_{k=1}  \left [   \frac{z_k^f}  {i \gamma\cdot q  +  m_k^f  }  +   \frac{\big (z_k^f\big )^{\!*} }  {i \gamma\cdot q  +  \big (m_k^f\big )^{\!*}  } \right  ]  ,
\end{equation}
where $m_k^f$ and $z_k^f$ are complex numbers. These parameters are fitted to the numerical  DSE solution for the quark propagator with $N=2$ on the real space-like axis 
$p^2\,\in[0,\infty)$. The set of parameters for each flavor of quarks are reported in Ref.~\cite{Serna:2020txe}.

With respect to the BSA defined by Eq.~\eqref{PS-BSA}, the scalar amplitudes $ \mathcal{\vec F} = (E_M,  F_M, G_M, H_M)$ are parametrized by a Nakanishi type of integral 
representation~\cite{Shi:2020pqe} that allows a simultaneously fit of even and odd components of the BSA under $k\cdot P \rightarrow - k\cdot P$ and therefore of equal-mass
as well as flavored mesons,
\begin{equation}
 \label{BSAPAR}
   \mathcal{F}_i(k,P)=\sum^{3}_{j=1}  \,  \int^1_{-1}  \!d\alpha\,  \rho_j (\alpha)  \frac{U_j \Lambda^{2n_j}}{(k^2+\alpha\, k\cdot P + \Lambda^2)^{n_j}} \, ,
\end{equation} 
where a dependence on the index $i$ is implicit for all parameters on the right-hand side.   The spectral densities $\rho_j (\alpha)$ are defined as,
\begin{align}
\rho_{1}(\alpha) &= \rho_{2}(\alpha)  = \frac{\Gamma(3/2)}{\sqrt{\pi}\,\Gamma(1)} \left[1+\sum^3_{n=1}\,\sigma_nC_n^{(1/2)}(\alpha)\right ],
\label{rho-ir} \\  
\rho_{3}(\alpha) &= \frac{3}{4}(1-\alpha^2),
\label{rho-uv}   
\end{align}
where $C_n^{(1/2)}(\alpha)$ are Gegenbauer polynomials of order 1/2. The scalar amplitude $H(k,P)$ is negligibly small and therefore neglected. The parameter sets of 
Eq.~\eqref{BSAPAR} are listed in the Appendix~\ref{app:BSAspar}. 

With the numerical values for the transverse-momentum dependent moments~\eqref{xm-LFWFs-01} at hand, one can reconstruct the meson's LFWF. In order to do so, 
we need an algebraic parametrization for $\Psi_{l_z}(x,\k^2_\perp )$ which allows us to reconstruct it by comparison of the calculated moments $ \langle x^m\rangle_{l_z} 
(\k^2_\perp )$~\eqref{xm-LFWFs-01} with the ones using the definition in Eq.~\eqref{defxm}. Motivated by the algebraic model described in Ref.~\cite{Almeida-Zamora:2023bqb}, 
we implement the  following parametrization for the LFWF:
\begin{eqnarray}
  \label{LFWFs-par}
  \Psi^{\rm rec.}_{l_z}(x,\k^2_\perp) = {\cal N}_M  \frac{\Delta^{2n}_M}{(\k^2_\perp+\Delta^2_M)^{n+1}}\, \phi_M(x,\alpha,\beta)\, ,
\end{eqnarray} 
where we fix $n=4$ for all mesons and define the function, 
\begin{eqnarray}
  \Delta^2_M(x)=\Lambda^2_1 + x(x-1)m^2_M + x(\Lambda^2_2-\Lambda^2_1)\, .
\end{eqnarray}
${\cal N}_M$, $\alpha$, $\beta$, $\Lambda_1$ and $\Lambda_2$ are unknown scale-dependent  parameters to be determined in the reconstruction process,
while $m_M$ is the ground-state  mass eigenvalue of the meson that corresponds to the BSA~\eqref{PS-BSA}.  For the Light-Front Distribution Amplitude (LFDA),
 $\phi_M(x,\alpha,\beta)$,  we employ the parametrization~\cite{Serna:2022yfp},
\begin{align}
\label{Psi-rec-L}
\phi_L(x,\alpha,\beta) &= x^\alpha   \bar x^\beta,   \\  \intertext{for light pseudoscalar mesons and}      
\phi_H(x,\alpha,\beta) & = 4x\bar x\,   e^{-4\,\alpha  x\bar x- \beta (x-\bar x)} \, ,
\label{Psi-rec-H}
\end{align}
for the heavy-light mesons and quarkonia with $\bar x = 1-x$. 

Finally, we reconstruct the LFWFs by minimizing the sum,
\begin{equation}
\label{LFWFs-rec}
 \epsilon\,  ( \mathcal{N}_M, \Lambda_1,\Lambda_2,\alpha, \beta) = \sum^{m_\textrm{max}}_{m=0}  \left | \frac{\langle x^m \rangle^\textrm{rec.}_{L,H}}{\langle x^m\rangle_{l_z}} - 1 \right | ,
\end{equation}
and where the moments $\langle x^m \rangle^\textrm{rec.}_{L,H}$ are calculated using Eq.~\eqref{defxm} with  Eq.~\eqref{Psi-rec-L}  or Eq.~\eqref{Psi-rec-H}, whereas 
$\langle x^m\rangle_{l_z}$ denotes the $\k^2_\perp$-dependent moments in Eq.~\eqref{xm-LFWFs-01}. We use $m_{\rm max} = 5$ moments in Eq.~\eqref{LFWFs-rec}, which 
allows for a robust reconstruction. While the $\k_\perp$-dependent part of the algebraic model only takes into account the dominant amplitude of the meson BSA~\cite{Almeida-Zamora:2023bqb}, 
we here assume that we can reconstruct the wave function with all scalar amplitudes of the BSA, the advantage being that in Eq.~\eqref{LFWFs-par} the relation between the LFWF
and the LFDA is simple. We verify that the LFWF of the light mesons and the heavy quarkonia can also be extracted directly from  Eq.~\eqref{xm-LFWFs-01}, that is without 
resorting to the above reconstruction, by performing the $dk^+dk^-$ integration with the help of a Feynman parametrization and a suitable variable transformation~\cite{Mezrag:2015mka}. 
However, in order to do so, a shift $k \rightarrow k - \eta P$ has to be applied  to the quark momenta in Eq.~\eqref{BSAwavefunc}, which becomes impractical for 
$P^2 \lesssim - m_D^2$, as the integration inevitably reaches poles in the quark propagators~\eqref{ccp}.

\begin{figure*}[t!]
\vspace*{-4mm}
\centering
\includegraphics[width=0.3\textwidth]{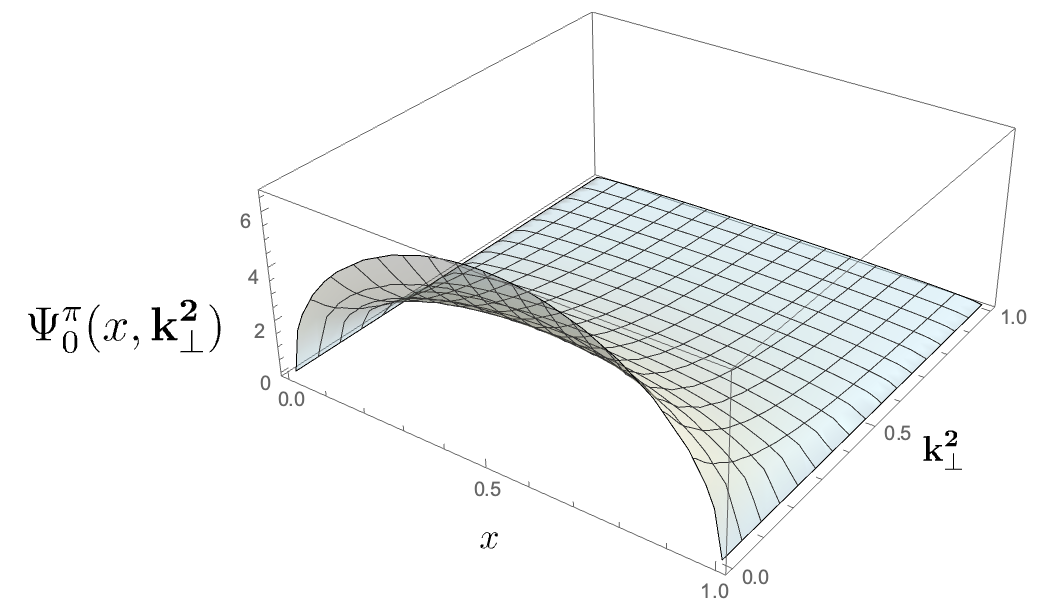}
\includegraphics[width=0.3\textwidth]{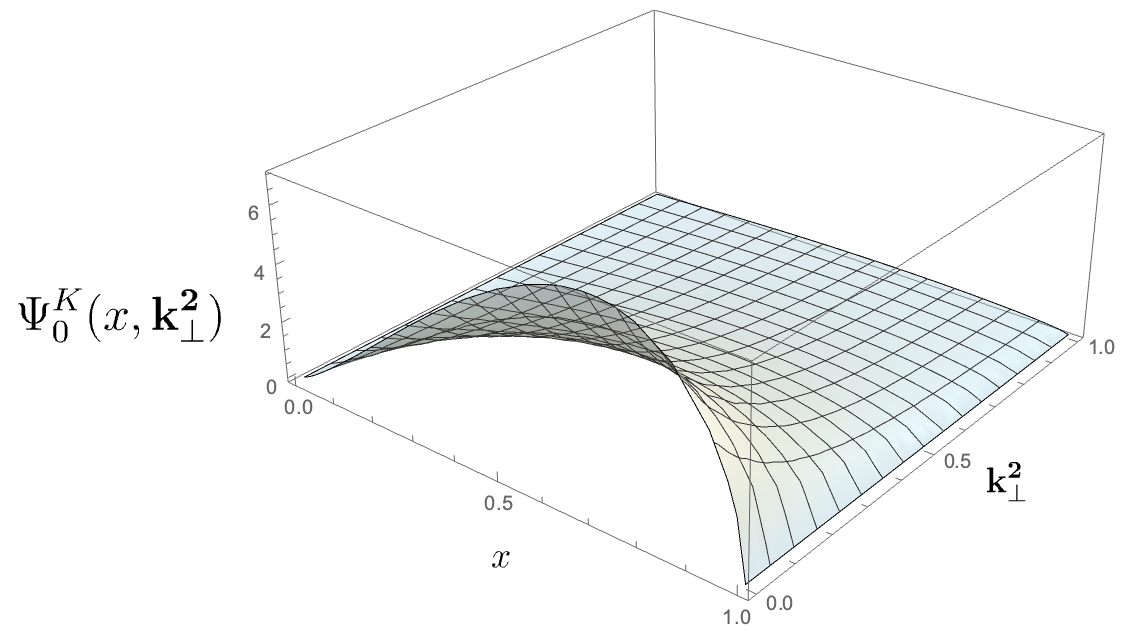}
\includegraphics[width=0.3\textwidth]{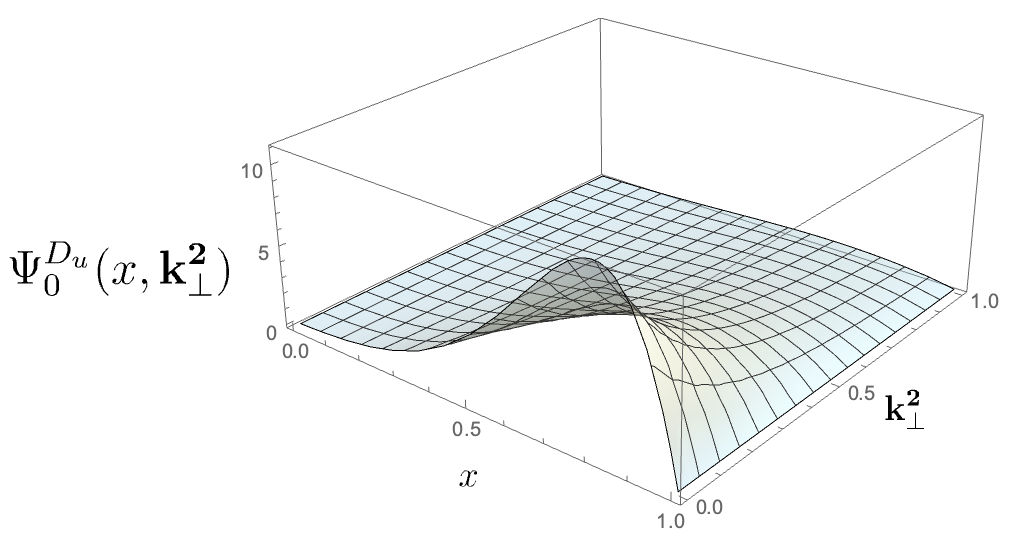}
\includegraphics[width=0.3\textwidth]{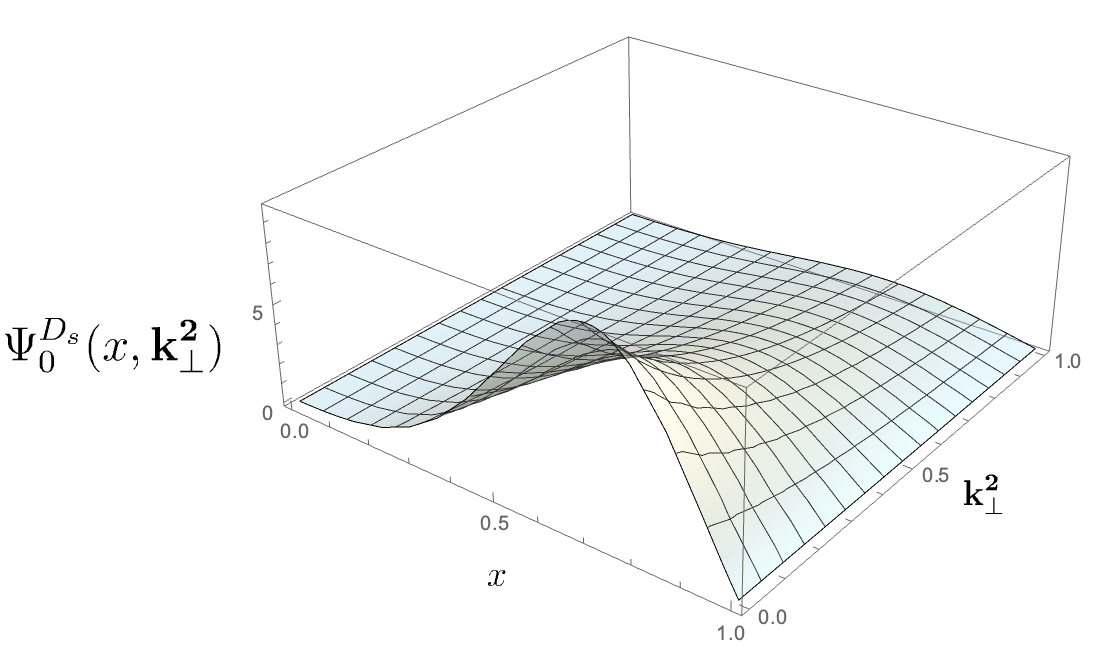}
\includegraphics[width=0.3\textwidth]{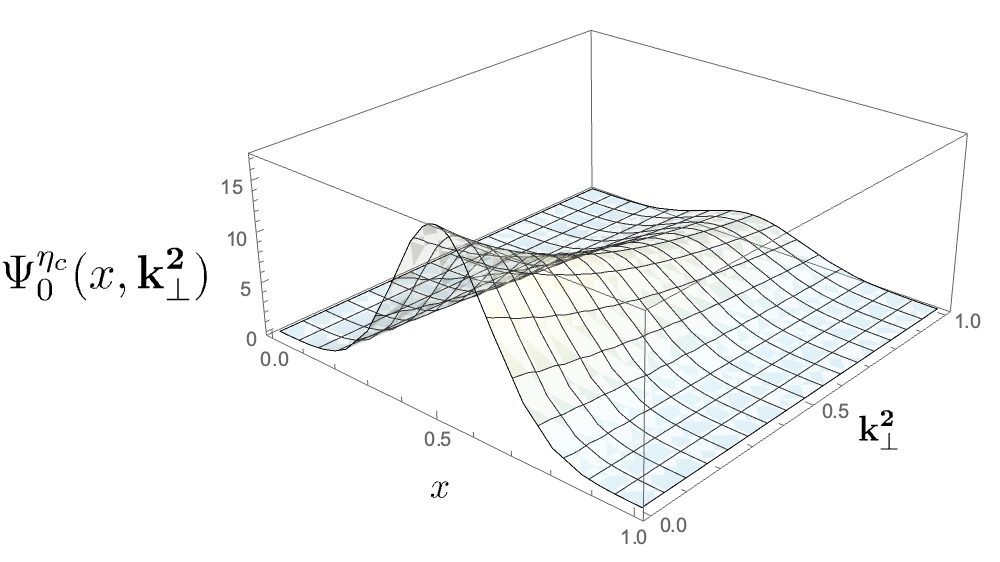}
\includegraphics[width=0.3\textwidth]{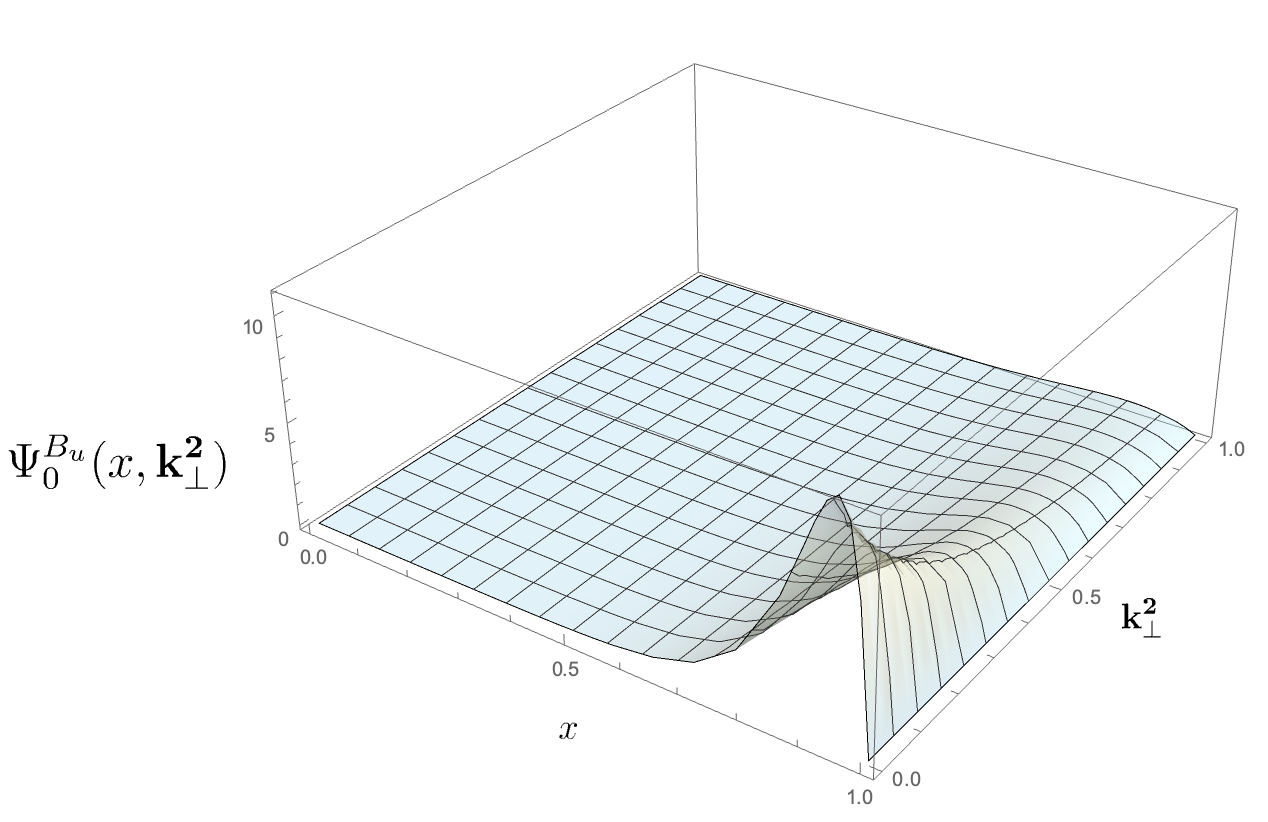}
\includegraphics[width=0.3\textwidth]{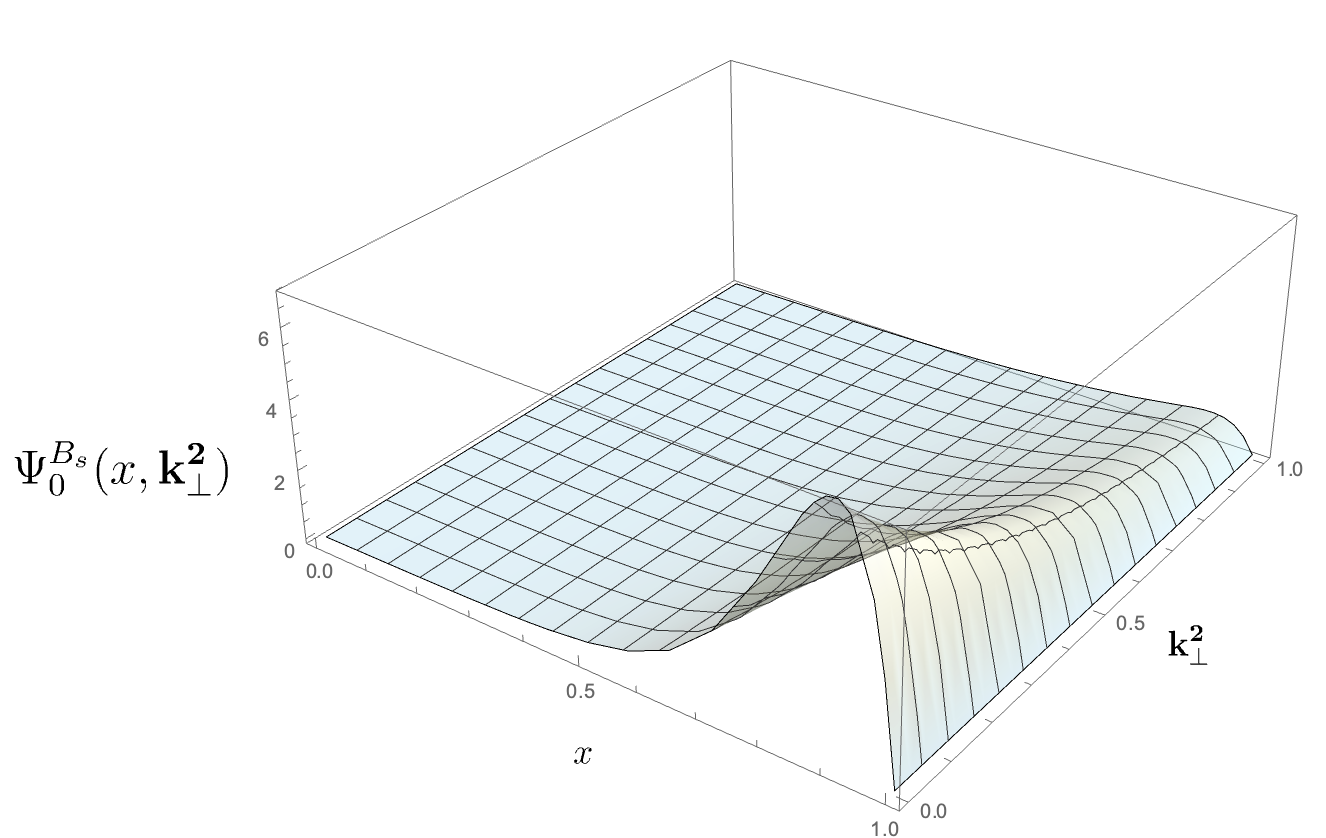}
\includegraphics[width=0.3\textwidth]{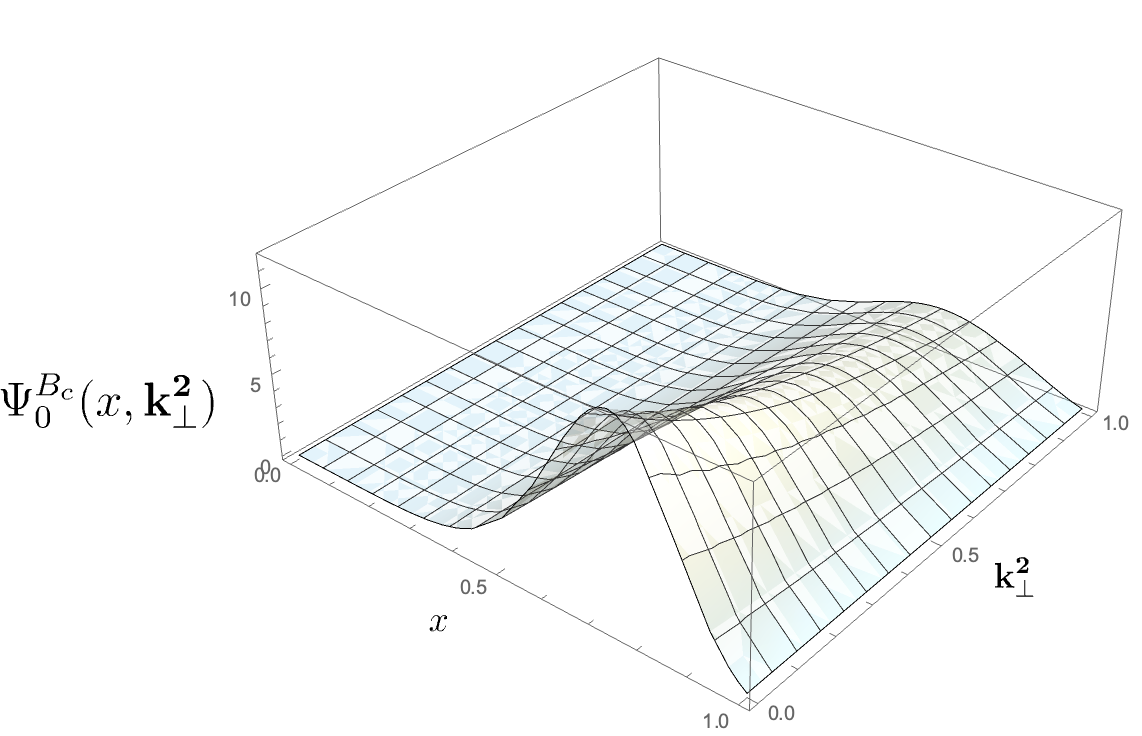}
\includegraphics[width=0.3\textwidth]{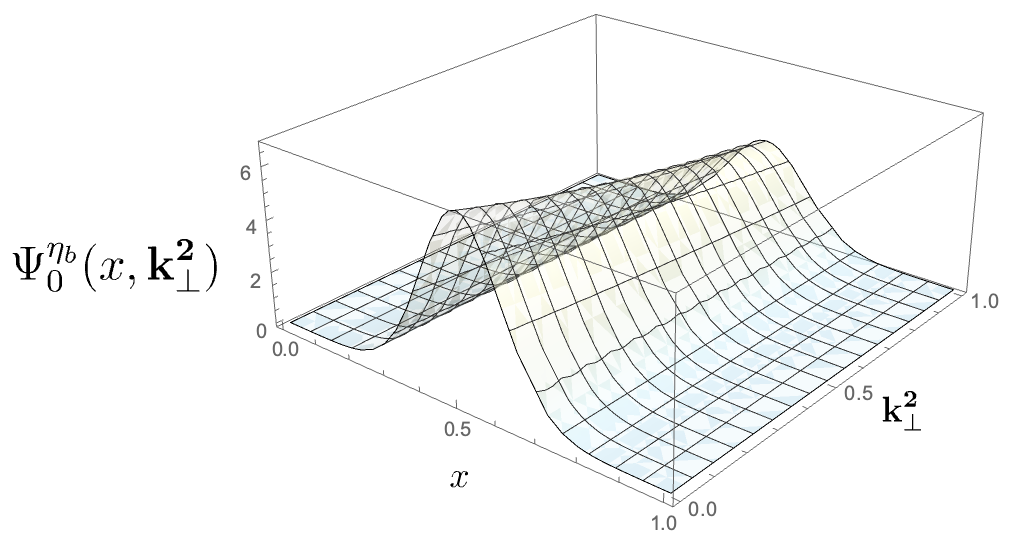}
\caption{Light-front wave functions of light and heavy pseudoscalar mesons for the angular projection  $l_{z}=0$ reconstructed from transverse-momentum dependent moments 
$\langle x^m\rangle_{l_z} (\k^2_\perp )$ at the scale $Q_0= 0.53\,$GeV. The units for the $\k^2_\perp$  axis are in $\text{GeV}^2$.} 
\label{Psifigs}
\end{figure*}

The fitted parameters are listed in Tab.~\ref{tab:Psi-par} for the angular projection $l_{z}=0$ and $l_{z}=1$.  Generally, the moments $\langle x^m\rangle_{l_z}$ are monotonously 
decreasing functions of $\k^2_\perp$ and tend to zero in the limit $\k^2_\perp \rightarrow \infty$.  They are, moreover,  suppressed with increasing values of $m$. 
The LFWFs of the mesons we consider are plotted in Fig.~\ref{Psifigs} for $l_z = 0$. For $| l_{z}| =1$ they  can be readily obtained from Eq.~\eqref{LFWFs-par} and 
Tab.~\ref{tab:Psi-par}. As expected,  $\Lambda_1=\Lambda_2$ and $\alpha=\beta$ for equal-mass mesons, whereas one has $\Lambda_1\neq\Lambda_2$ and $\alpha\neq\beta$
for flavored mesons.  In case of the light mesons, we observe that at  large $\k_\perp^2$ the LFWFs decay as $\psi_0\left(x, \k_\perp^2 \right ) \sim 1 /\k_\perp^2$  
and $\psi_1\left(x, \k_\perp^2\right) \sim 1/ \k_\perp^4$,  in agreement with perturbative QCD~\cite{Ji:2003fw}, and the wave functions reflect the dynamical chiral symmetry breaking properties 
of the Bethe-Salpeter wave functions. That is, their $x$-dependence is broad at low $\bm{k}_\perp^2$  and narrows as  $\bm{k}_\perp^2$  increases, approaching the asymptotic form $x(1-x)$. 
The effect of SU(3) flavor symmetry breaking  is clearly visible for the kaon, as the heavier strange quark gains more support at larger $x$ and the LFWF is skewed. 
This observation is exacerbated for charm and bottom mesons with a strong shift to larger $x$ of the heavy quark's support, whereas a ridge forms along $\k_\perp^2$ 
with increasing  meson mass. Their LFWFs decay at larger $\k_\perp^2$, in particular those of the $\eta_c$ and $\eta_b$ quarkonia. 

So far, we have not discussed the scale dependence of the LFWF. Since we use a model for the gluon interaction~\cite{Serna:2020txe,daSilveira:2022pte} in ladder truncation, the 
renormalization scale of the quark propagators $S_f(k)$ is not that in a given scheme in perturbative QCD. We therefore determine the scale with the experimental value of the the 
momentum fraction carried by the valence quarks from a $\pi N$ Drell-Yan analysis, $2 \langle x \rangle_v = 0.47(2)$  at a scale of $Q^2 = 4\,$GeV$^2$~\cite{Sutton:1991ay,Gluck:1999xe}. 
By solving the NLO DGLAP evolution equations numerically with \texttt{QCDNUM}~\cite{Botje:2010ay} we determine our model scale to be $Q_0 = 0.53\,$GeV.

\section{Distribution functions}

With the meson's LFWF at hand one can readily derive three distributions. The leading-twist distribution amplitude of the meson is obtained by integrating over the quark's transverse 
momentum~\cite{Brodsky:1997de}, 
\begin{equation}
    \phi_M(x) = \frac{1}{16\pi^3} \int^{\Lambda^2_\perp} \! \Psi_{0}(x,  \k^2_\perp ) \, d \k_\perp \, , 
\end{equation}
normalized with weak decay constant $f_M$,
\begin{equation}
    \int^1_0 \! \phi_M(x ) \,  dx = f_M \, . 
\end{equation}
The leading-twist time-reversal even TMD is given in terms of the minimal Fock-state LFWFs as~\cite{Pasquini:2014ppa,dePaula:2022pcb},
\begin{align}
  \label{TMD-def}
     f_M (x, \k^2_\perp ) =   \,  \frac{1}{(2 \pi)^{3}} & \Big ( \left|\Psi_{0}(x,  \k^2_\perp) \right |^2  
     + \,  \k^2_\perp  \left |  \Psi_{1}(x, \k^2_\perp )\right|^2\Big )  , 
\end{align}
and the PDF is obtained therefrom by integrating over the quark's transverse momentum:
\begin{equation}
    q_M (x,Q_0 ) =  \int^{\Lambda^2_\perp} \!  f_{M} (x,  \k^2_\perp ) \, d \k_\perp \,.
  \label{PDFdef}  
\end{equation}
The normalization condition that guarantees baryon number conservation of the LFWF is:
\begin{equation}
   \int^1_0 q_{M}   (x,Q_0 )\, dx = 1 \, . 
\end{equation}

\begin{figure*}[t!]
\vspace*{-2mm}
\centering
\includegraphics[width=0.3\textwidth]{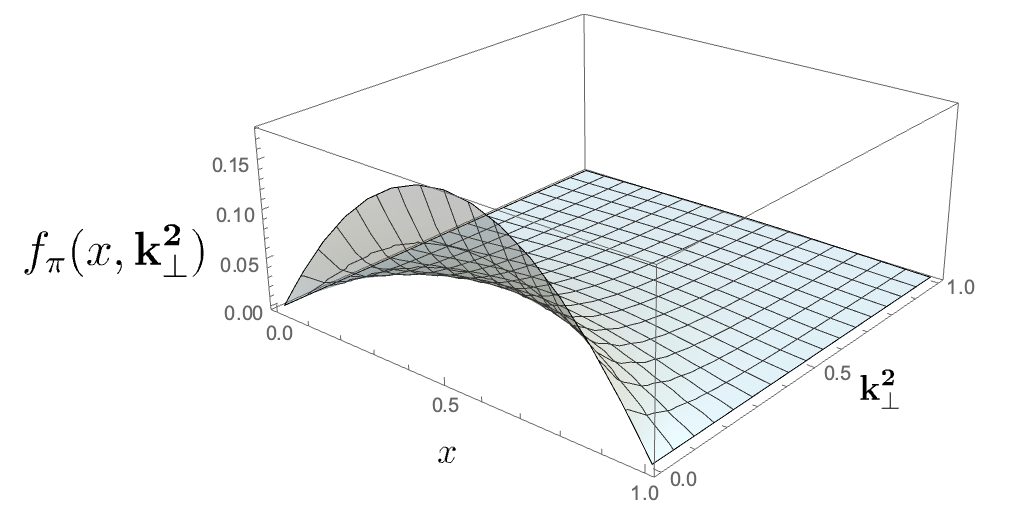}
\includegraphics[width=0.3\textwidth]{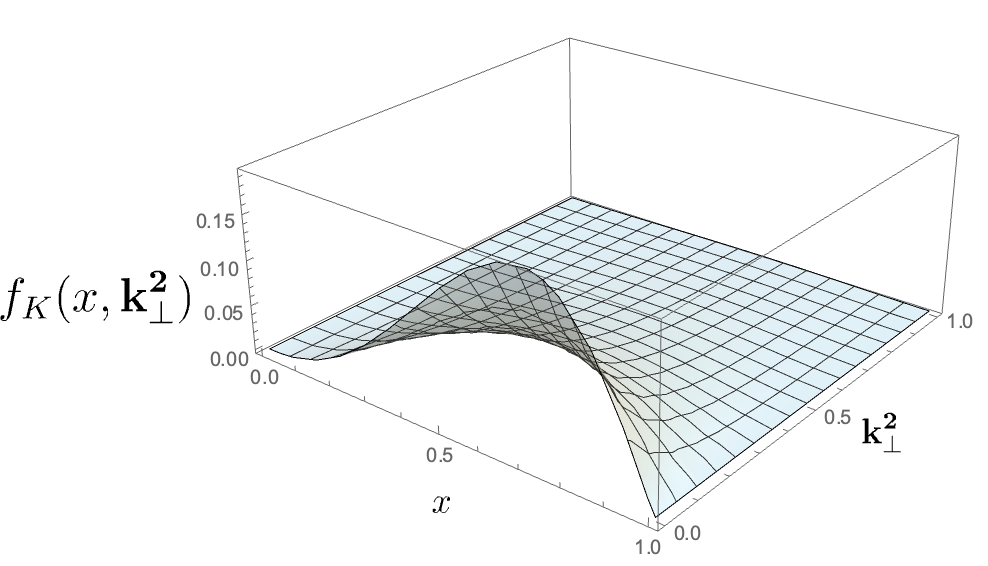}
\includegraphics[width=0.3\textwidth]{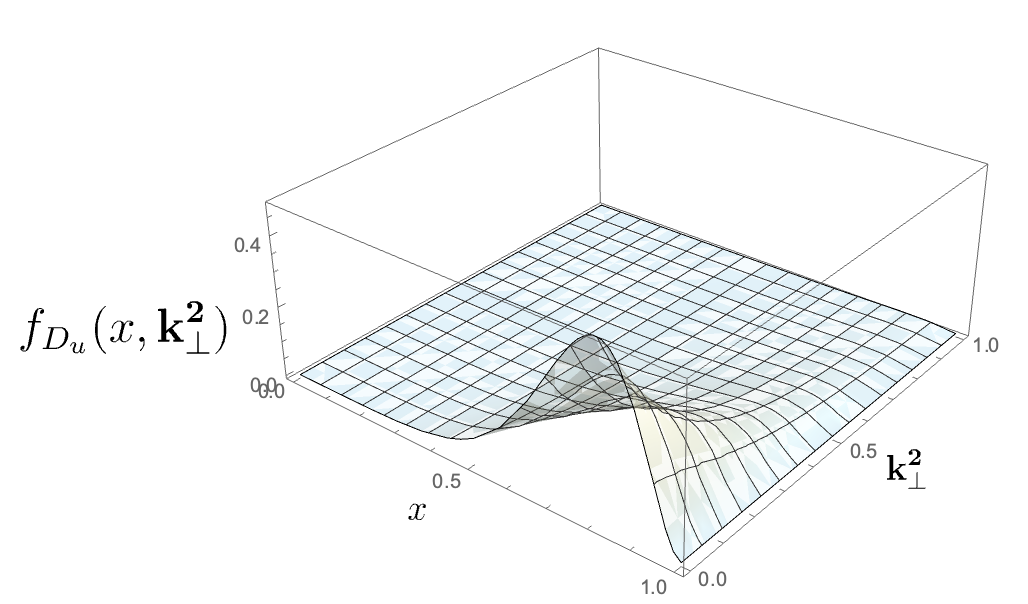}
\includegraphics[width=0.3\textwidth]{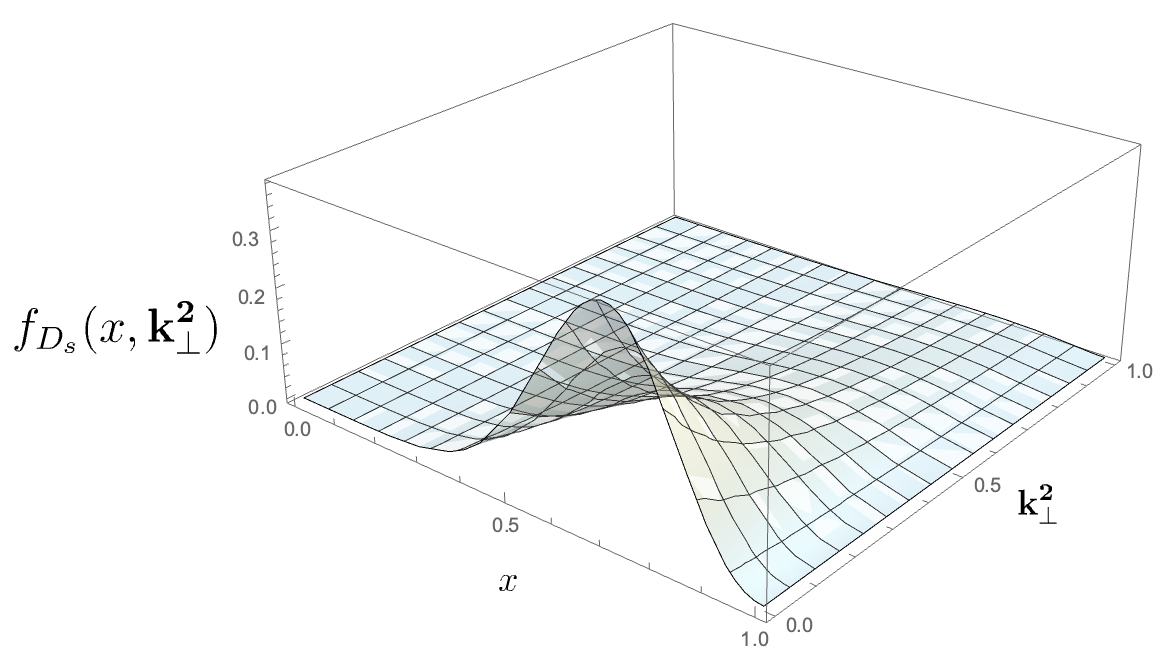}
\includegraphics[width=0.3\textwidth]{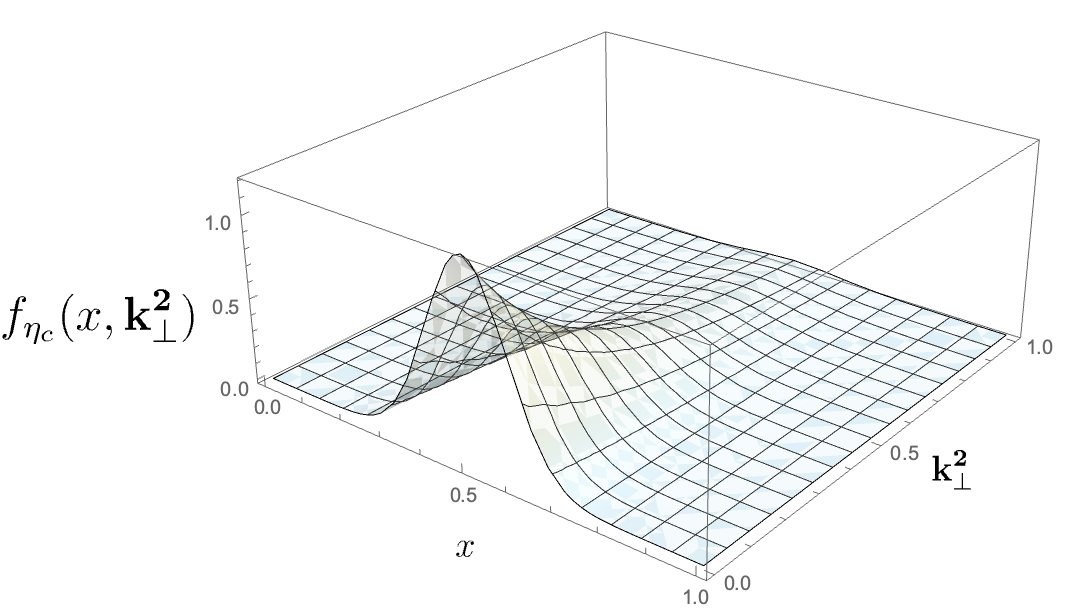}
\includegraphics[width=0.3\textwidth]{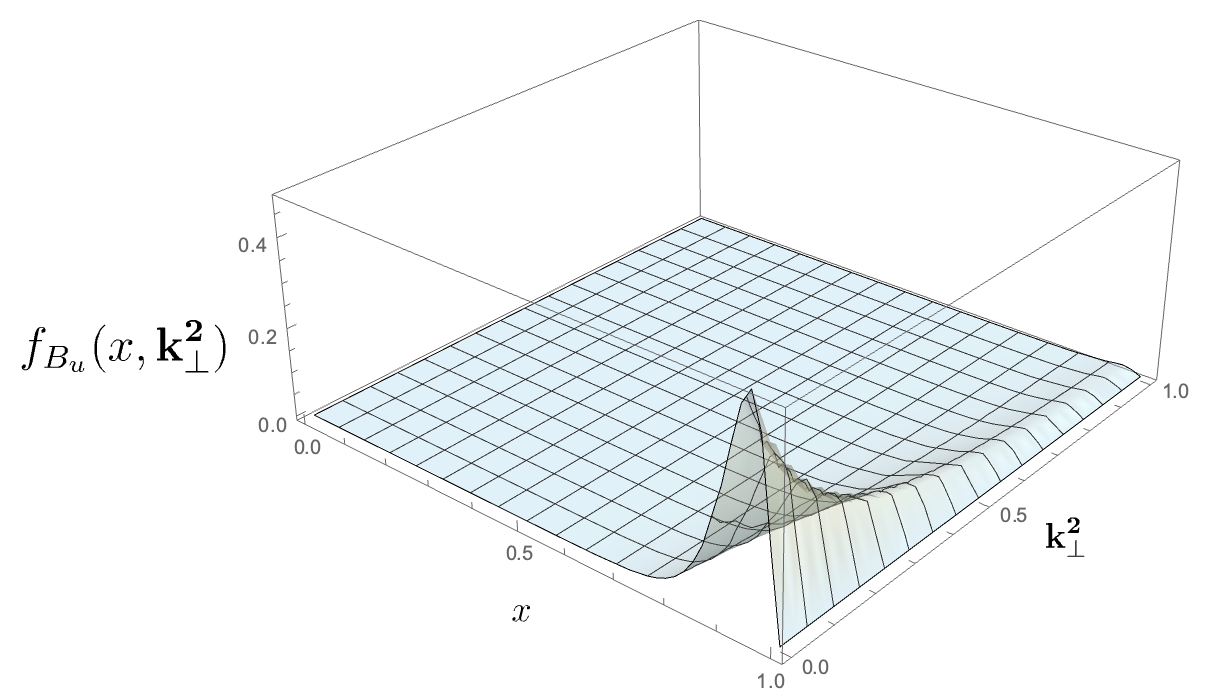}
\includegraphics[width=0.3\textwidth]{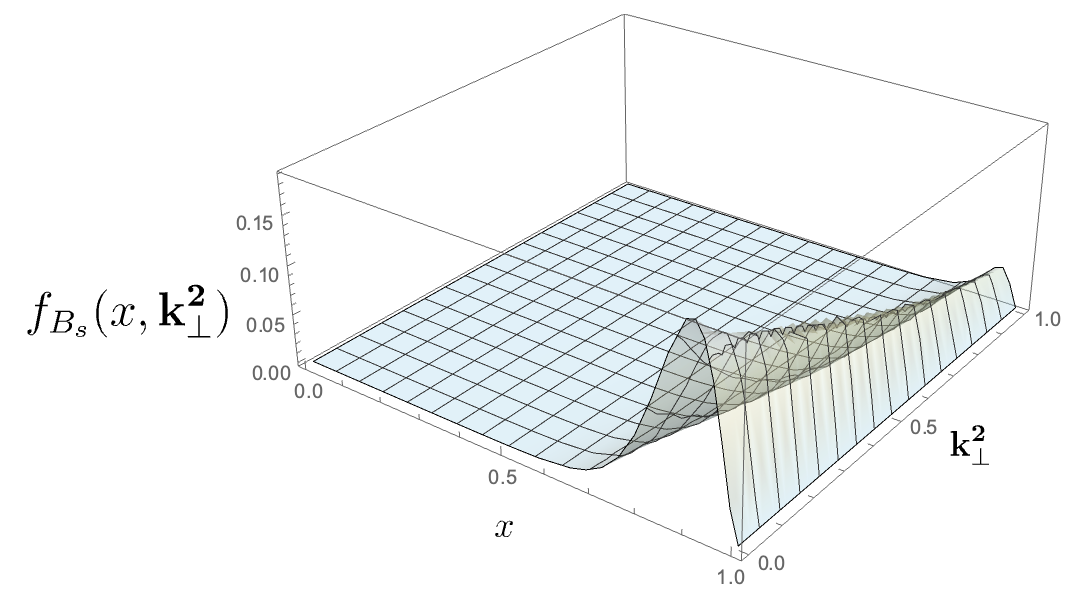}
\includegraphics[width=0.3\textwidth]{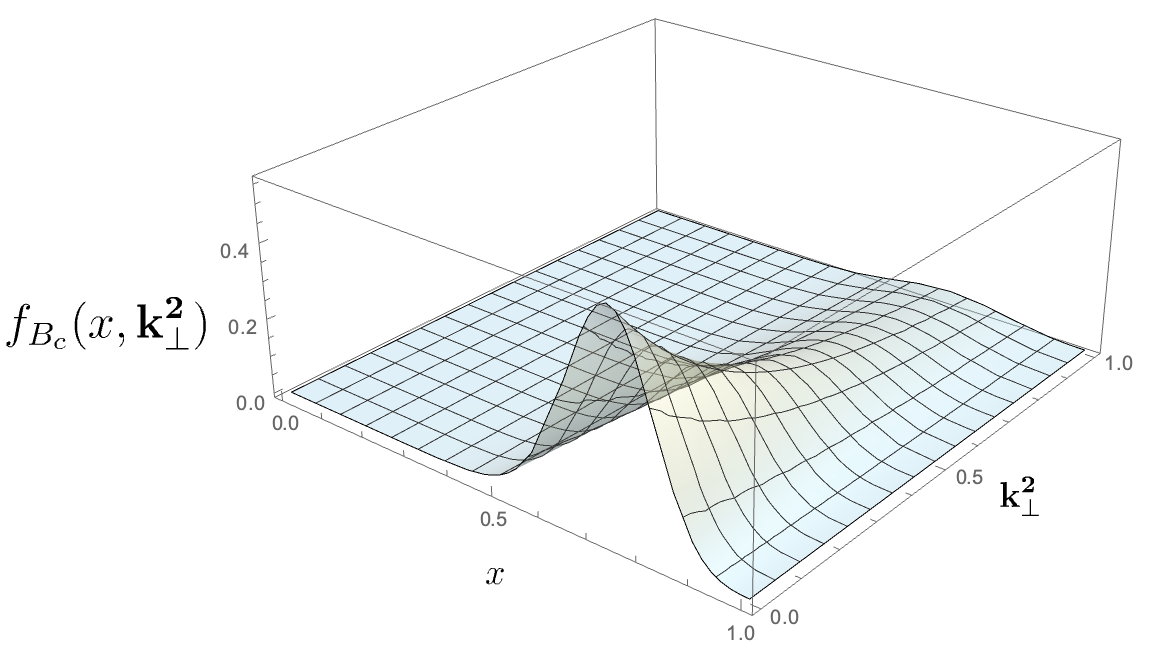}
\includegraphics[width=0.3\textwidth]{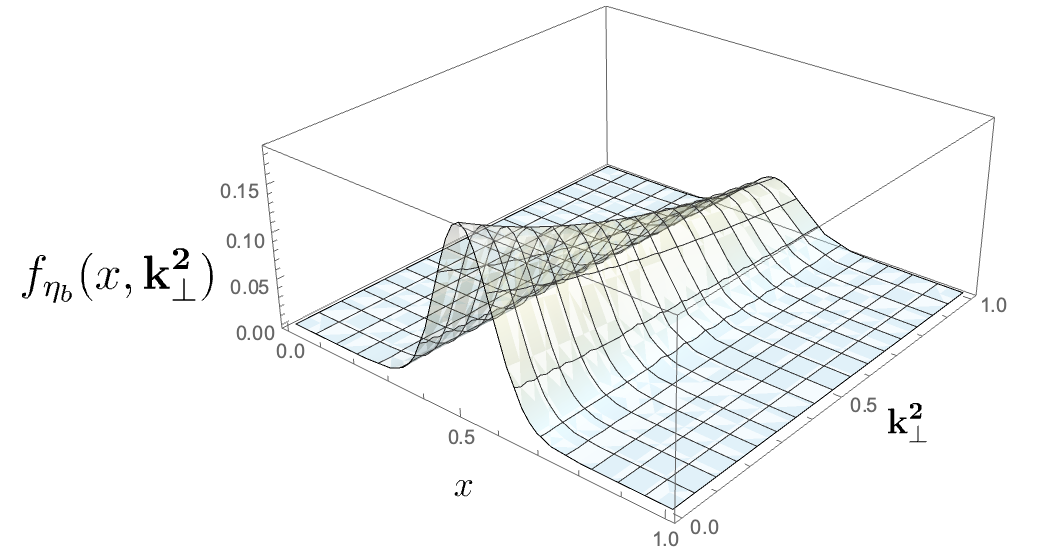}
\caption{Transverse momentum distributions of pseudoscalar mesons for the minimal Fock-state configuration in Eq.~\eqref{TMD-def} at the scale $Q_0 = 0.53\,$GeV. 
The units for the $\k^2_\perp$  axis are in $\text{GeV}^2$.}
\label{TMDfigs}
\end{figure*}

\begin{figure*}[t!]
\vspace*{3mm}
\includegraphics[width=0.46\textwidth]{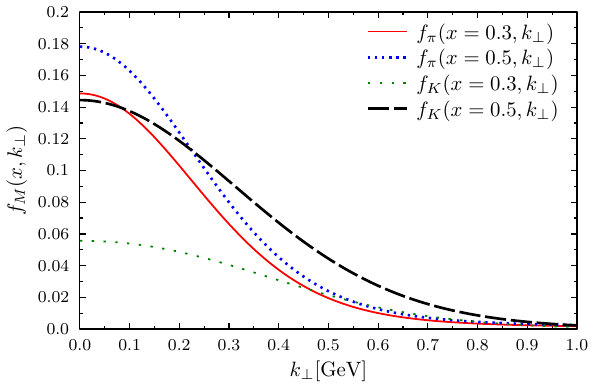}
\includegraphics[width=0.46\textwidth]{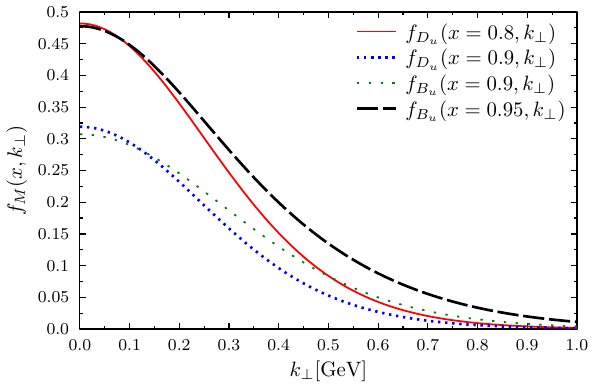}
\caption{TMD of light and heavy mesons as a function of $k_\perp$  for selected values of $x$. An inflection point from a concave to a convex function generally occurs at about 
$k_\perp \approx 0.3$\,GeV. }
\label{inflectpoint}
\end{figure*}

\begin{figure*}[t!]
\centering
\includegraphics[width=0.49\textwidth]{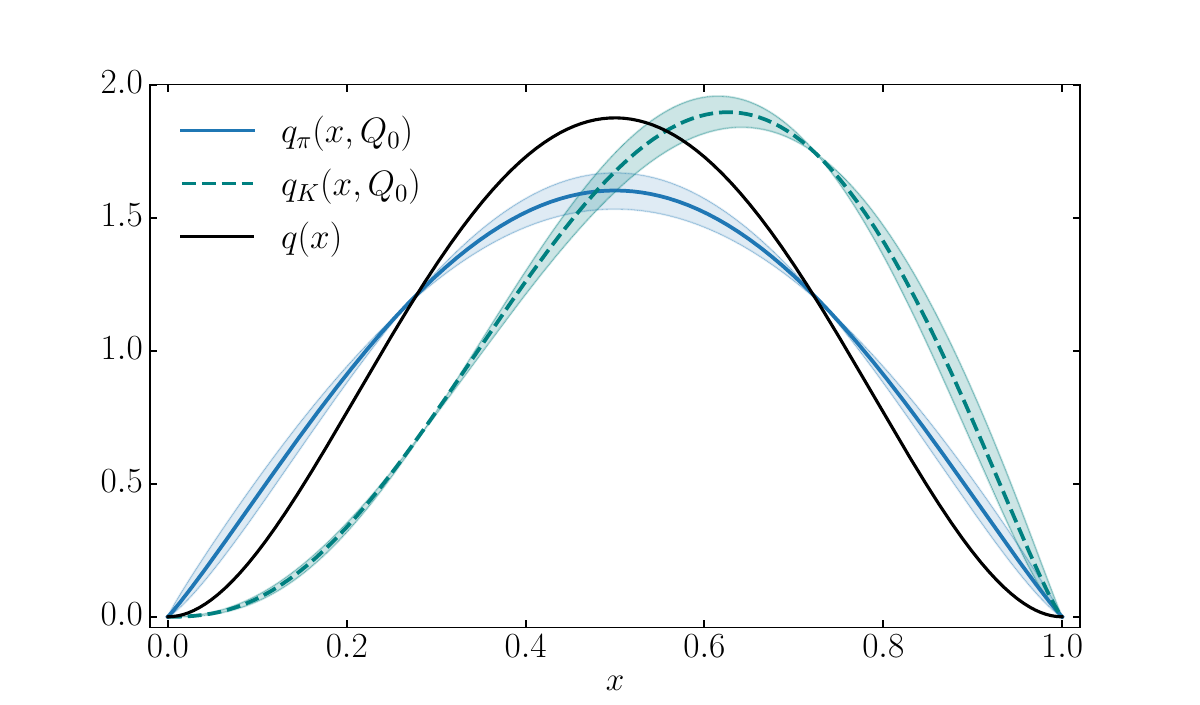}
\includegraphics[width=0.49\textwidth]{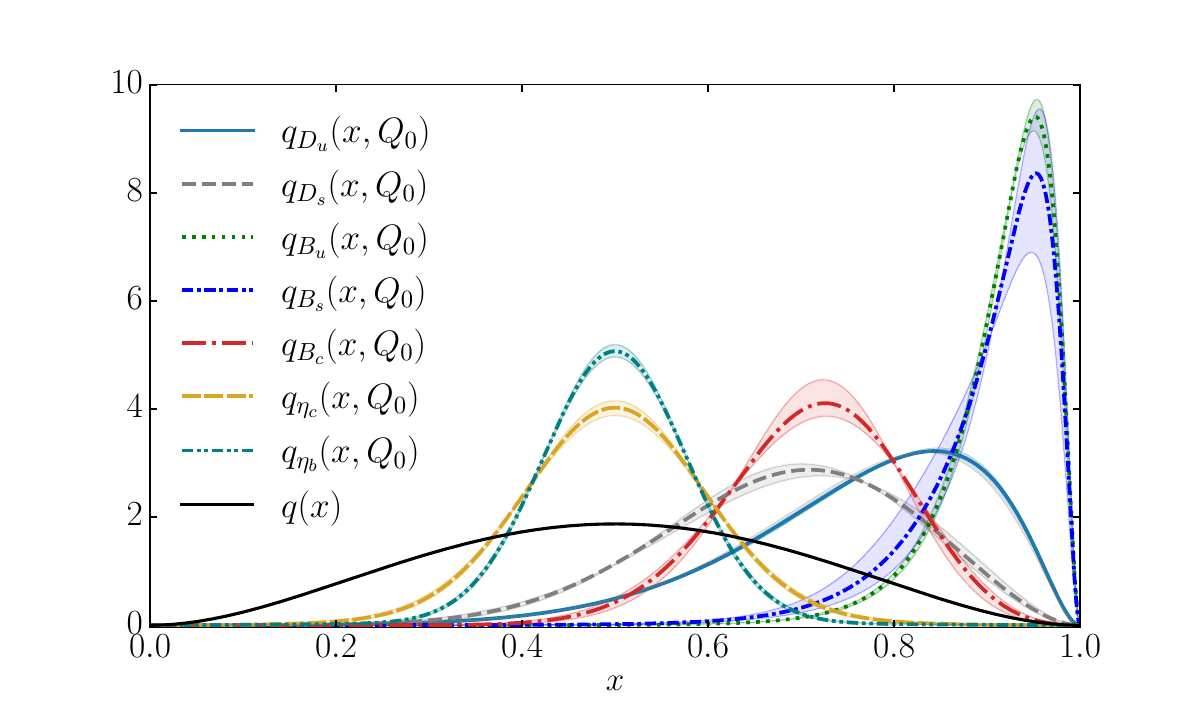}
\caption{Parton distribution functions of light and heavy mesons calculated at the scale $Q_0 = 0.53\,$GeV. For comparison, the asymptotic form $q (x) \sim [x(1-x)]^2$ is plotted. 
The error bands are due to the reconstruction procedure and stem from the uncertainties in the parameters ${\cal N}_M$, $\alpha$, $\beta$, $\Lambda_1$ and $\Lambda_2$ in 
Eq.~\eqref{LFWFs-par}. }
\label{PDFfigs}
\end{figure*}

In Fig.~\ref{TMDfigs}, the unpolarized TMDs of the pseudoscalar mesons are presented as functions of $\k_\perp^2$ and $x$. We point out that this includes the first TMD 
calculation of $D$ and $B$ mesons within the DSE-BSE approach. The transverse distributions mirror the functional behavior of the LFWF discussed in Sec.~\ref{LFWFsec}. 
They are smooth functions that decay with increasing  $\k^2_\perp$ and as functions of $x$. As illustrated in Fig.~\ref{inflectpoint}, an inflection point from a concave to 
a convex function of $k_\perp$ is located at about 300~MeV in the case of the pion and kaon, as was already observed in Ref.~\cite{Shi:2020pqe}. It persist at the same scale 
for the $D$, $B$, $\eta_c$ and $\eta_b$. However, the falling off with $\k_\perp^2$ of the TMD of these mesons is significantly attenuated and their $x$-dependence is narrow. The
latter observation is consistent with the fact that dynamical chiral symmetry breaking is less important in heavy mesons. We remind that the $x$-dependence of the
TMDs in Eq.~\eqref{LFWFs-par} does not factorize.

We conclude this section with the presentation of the PDFs in Fig.~\ref{PDFfigs}, which we obtain from the TMDs via Eq.~\eqref{PDFdef} with the value $\Lambda^2_\perp=8\,\text{GeV}^2$.
We stress that our results remain unchanged for $\Lambda^2_\perp\geq8\,\text{GeV}^2$. The PDFs presented in Fig.~\ref{PDFfigs} can be fitted with the Gegenbauer expansion,
\begin{equation}
   q (x , Q_0) = 30 \left [x(1-x) \right ]^2 \Bigg [1+\sum_{j=1}^{j_m} a_j C_j^{5 / 2}(2 x-1) \Bigg ] .
 \label{PDFfit}  
\end{equation}
We find that all PDFs can be reproduced with $j_m=17$. For mesons composed of the same valence quarks, only even Gegenbauer polynomials contribute to the parameterization. 
On the left-hand side of Fig.~\ref{PDFfigs}, the PDFs of the pion and kaon are compared to the asymptotic form of Eq.~\eqref{PDFfit}, $q (x ,Q_0) = 30[x(1-x)]^2$, in the limit $Q_0 \rightarrow \infty$.
We note again a clear broadening of the distributions which is nothing but a manifestation of dynamical chiral symmetry breaking in the meson's Bethe-Salpeter wave function. 
Likewise, SU(3) breaking is apparent and the strange quark carries a larger fraction of the kaon's longitudinal light-front momentum. On the right-hand side of Fig.~\ref{PDFfigs}, 
the PDFs of heavy quarkonia and heavy-light mesons are shown. Their functional behavior is analogous to that observed for their distribution amplitudes, see for example
Refs.~\cite{Serna:2020txe,Serna:2022yfp}: with increasing quark masses, the distribution functions become narrower and in case of an infinitely heavy quarkonia, one
expects $q (x) \simeq \delta (x-1/2)$.  The support of the $D$-meson distributions is shifted to larger $x$, even more so for the $B$ mesons as is clear from comparison with the 
asymptotic distribution $q(x)$. 

\begin{figure}[t!]
\vspace*{-5mm}
\hspace*{-3mm}
\includegraphics[width=0.53\textwidth]{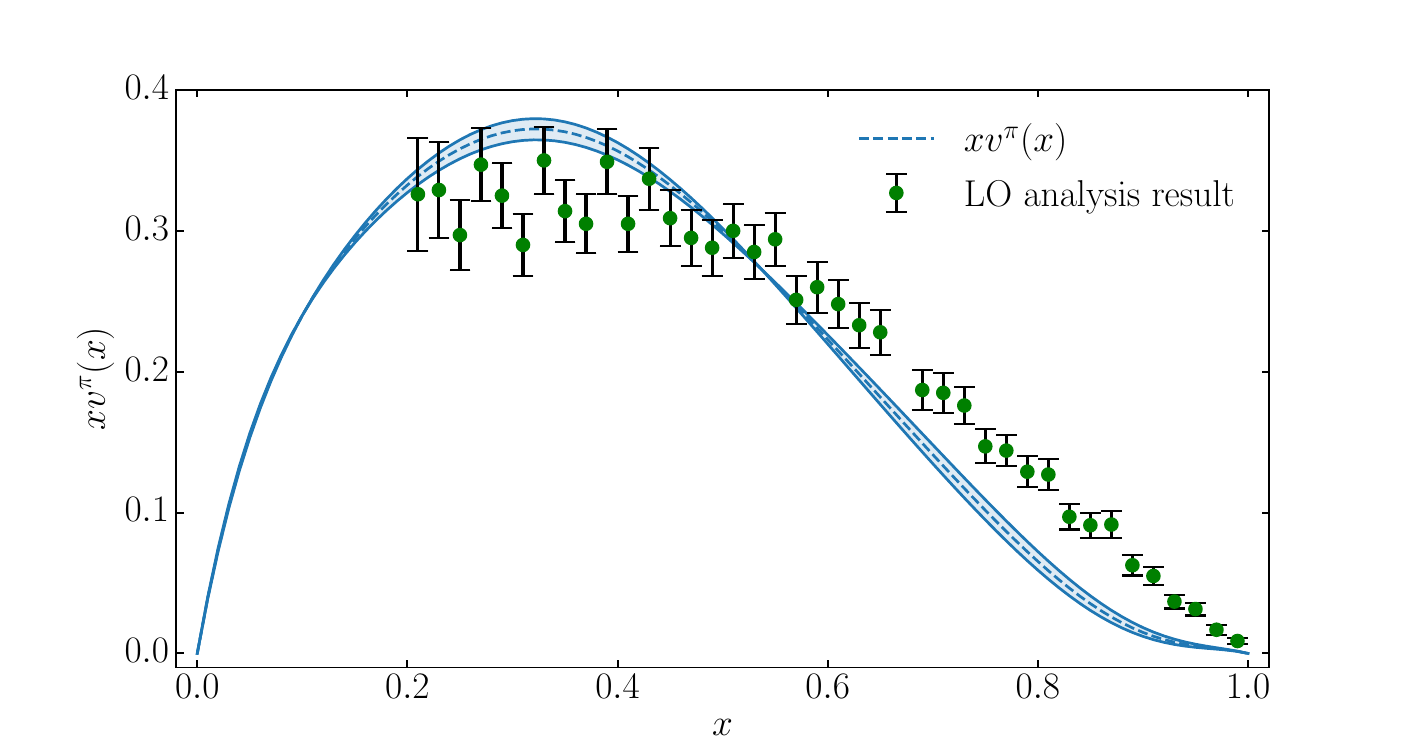}
\caption{The pion valence PDF evolved to  $Q=4$ GeV and compared with a LO analysis result~\cite{Conway:1989fs}. }
\label{fig:evolved-PDF}
\end{figure}

In Fig.~\ref{fig:evolved-PDF} we show our results for  the pion valence distribution $v^\pi$  (multiplied by Bjorken $x$) as a function of $x$. We evolve our results at 
$Q_0=0.53\,$GeV to a scale of $Q=4\,$GeV using the NLO DGLAP equations of the QCD evolution program  \texttt{QCDNUM}~\cite{Botje:2010ay} to compare with 
the LO analysis result of Ref.~\cite{Conway:1989fs}. Finally, the valence distribution function can be very accurately parametrized with a Chebyshev expansion, as follows:
\begin{equation}
  x v^\pi = N_v\, x(1-x)\left[1+\sum_{n=1}^{6} b_nU_n(2 x-1) \right ] .
\end{equation}


\section{Final remarks}

We projected the Bethe-Salpeter wave function of pseudoscalar mesons on the light front by computing $\k^2_\perp$-dependent moments of the leading Fock-state LFWF,
where the wave functions employed are computed in an improved ladder kernel that accounts for the flavor difference of the light and heavy quarks~\cite{Serna:2020txe}.
The LFWF~\eqref{LFWFs-01} and TMD~\eqref{TMD-def} of pseudoscalar mesons we obtain provide an insightful picture of their longitudinal momentum fraction and 
transverse momentum dependences as a function of the quark-antiquark mass difference. 

We reproduce earlier DSE-BSE calculations of the TMDs for the pion and kaon~\cite{Shi:2018zqd,Shi:2023oll} and for heavy quarkonia~\cite{Kou:2023ady} and report
the first TMD and PDF of $D, D_s, B, B_s$ and $B_c$ mesons using a functional approach. Not surprisingly, and mirroring the behavior of the distribution 
amplitudes~\cite{Serna:2020txe,Serna:2022yfp}, the larger the quark-mass difference, the narrower the LFWF, TMD and PDF become and the more they are skewed towards 
larger $x$. Moreover, with increasing meson mass, the TMDs fall off much slower than those of the pion and kaon as a function of  $\k^2_\perp$. One therefore ought to be aware 
of this behavior when studying the transverse-momentum dependence heavy meson production amplitudes. Finally, we determine our model scale by evolving the pion valence 
quark distribution function to experimental scale.  \\


\acknowledgments

We would like to acknowledge Chao Shi for insightful discussions and for providing the PDF evolution code. F.E.S. gratefully acknowledges the EIC Theory Institute at BNL for their financial
 support during his research stay.  B.\,E. and G.\,K. participate in the Brazilian network project \emph{INCT-F\'isica Nuclear e Aplica\c{c}\~oes\/}, grant no.~464898/2014-5. This work was 
 supported by the S\~ao Paulo Research Foundation (FAPESP), grant no.~2023/00195-8 (B.~E.) and no.~2018/25225-9 (G.~K.), and by the National Council for Scientific and Technological 
 Development (CNPq), grant no.~409032/2023-9 (B.~E.) and no.~309262/2019-4 (G.~K.).  


\appendix

\section{Bethe-Salpeter Amplitude Parameters}\label{app:BSAspar}

In Table~\ref{Table:parameters} we list all the parameters associated with BSA of the pseudoscalar mesons. The corresponding parametrization is detailed in Eq.~\eqref{BSAPAR}.

\begin{table}[t!]
\caption{Parameters of the spectral representation of the meson's BSA~\eqref{BSAPAR}. (Dimensioned quantities in GeV).
\label{Table:parameters}
}
\begin{center}

\begin{tabular*}
{\hsize}
{
l@{\extracolsep{0ptplus1fil}}
l@{\extracolsep{0ptplus1fil}}
 r@{\extracolsep{0ptplus1fil}}
r@{\extracolsep{0ptplus1fil}}
r@{\extracolsep{0ptplus1fil}\hspace{0.2cm}}
r@{\extracolsep{0ptplus1fil}}
r@{\extracolsep{0ptplus1fil}}
r@{\extracolsep{0ptplus1fil}}
r@{\extracolsep{0ptplus1fil}}
r@{\extracolsep{0ptplus1fil}}
r@{\extracolsep{0ptplus1fil}}}\hline\hline
   &$\Lambda$  & $U_1$ & $U_2$ & $U_3$ &$n_1$ &$n_2$ &$n_3$ & $\sigma_1$ & $\sigma_2$ & $\sigma_3$\\[0.0ex]\hline
 $E_\pi$ & 1.42 & 0.75 & 1.45 & $-1.21$
    & 4 & 5&6&0.00&1.13&0.00 \\
   $F_\pi$ & 1.43 & $-$0.44 &0.94&0.01
    & 3 & 4&1&0.00&$ $-1.30 &0.00 \\
    $G_\pi$ & 0.80 & $-$0.05 &0.04&0.01
    & 4 & 5&1&38.00&2.81 &1.25 \\\hline  
 $E_K$ & 1.33 & 2.36 & $-$1.09 & $-$0.27
    & 4 & 5&6&$-$0.97&1.28&$-$1.53\\  
 $F_K$ & 1.56 & $-$0.39 & 0.80 & 0.01
    & 3 & 4&1&$-$0.64&$-$1.09&1.20\\             
 $G_K$ & 1.57 & 0.00 & 0.25& $-$0.25
    & 5 & 6 &5&$-$1.64&2.34&$-$3.18\\ \hline  
 $E_{\eta_c}$ & 1.84 & 2.08 & $-$1.13 & 0.06
    & 4 & 5&1&0.0&$-$1.65&0.00\\     
 $F_{\eta_c}$ & 2.55 & $-$0.02 & 0.15 & 0.00
    & 4 & 6&1&0.0&$-$2.14&0.00\\      
 $G_{\eta_c}$ & 2.71 & 0.82 & $-$0.08 & $-$1.12
    & 8 & 12&8&$-$0.33&0.00&0.75\\ \hline         
 $E_{\eta_b}$ & 3.31 & 0.16 & 0.78 & 0.06
    & 4 & 5&1&0.00&$-$2.22&0.00\\   
 $F_{\eta_b}$ & 3.50 & 0.00 & 0.05 & 0.00
    & 4 & 6&1&0.00&$-$2.30&0.00\\         
 $G_{\eta_b}$ & 3.86 & 0.08& $-$0.04 & $-$1.54
    & 8 & 12&8&$-$1.00&0.00&3.08\\ \hline    
 $E_{D_u}$ & 1.44 & 2.49 & $-$1.50 & 0.01
    & 4 & 5&1&0.133&$-$1.48&0.57\\       
 $F_{D_u}$ & 2.03 & $-$0.07 & 0.278 & 0.00
    & 4 & 6&1&$-$0.41&$-$1.92&0.90\\      
 $G_{D_u}$ & $-$2.60 & 0.14 & 0.20 & $-$0.36
    & 8& 12&8&$-$1.09&$-$1.63&2.23\\ \hline 
 $E_{D_s}$ & 1.56 & 2.44 & $-$1.45 & 0.01
    & 4 & 5&1&0.23&$-$1.40&0.15\\  
 $F_{D_s}$ & 1.47 & 0.33 & $-$0.12 & 0.00
    & 4 & 6&1&0.30&$-$1.88&$-$0.72\\       
 $G_{D_s}$ & $-$1.50 & $-$0.19 & 0.12 & 0.00
    & 4 & 5&1&3.25&0.95&$-$5.07\\  \hline        
 $E_{B_u}$ & 1.55 & 2.61 & $-$1.60 & $-$0.03
    & 4 & 5&1&0.13&$-$2.29&$-$0.28\\       
 $F_{B_u}$ & 2.36 & $-$0.05 & 0.14 & 0.00
    & 4 & 6&1&0.34&$-$2.30&$-$1.04\\       
 $G_{B_u}$ & 2.16 & 0.142 & $-$0.01 & $-$0.21
    & 5 & 6&4&$-$1.36&$-$0.98&4.26\\  \hline      
 $E_{B_s}$ & 1.71 &  2.36 & $-$1.31 & $-$0.05
    & 4 & 5&1&0.55&$-$2.11&$-$1.55\\       
 $F_{B_s}$ & 2.21 &  $-$0.01 & 0.10 & 0.00
    & 4 & 6&1&0.60&$-$2.17&$-$1.83\\       
  $G_{B_s}$ & 1.58 &   $-$0.41 & 0.26 & 0.00
    & 4 & 5&1& 1.35&$-$1.57&$-$3.80\\  \hline      
 $E_{B_c}$ & 2.14 &  1.94 & $-$0.96 & 0.02
    & 4 & 5&1& 0.33&$-$2.13&$-$ 0.88\\      
 $F_{B_c}$ & 2.78 &  0.00 & 0.07& 0.00
    & 4 & 6&1&  0.33&$-$2.25&$-$0.99\\      
  $G_{B_c}$ & 2.45 &  0.01 & 0.00& 0.05
    & 5 & 6&4&$-$43.57&4.04&127.57\\   \hline  \hline                   
\end{tabular*}
\end{center}
\end{table}



\begin{thebibliography}{99}

\bibitem{Brodsky:1997de}
S.~J.~Brodsky, H.~C.~Pauli and S.~S.~Pinsky,
Phys. Rept. \textbf{301}, 299-486 (1998)
doi:10.1016/S0370-1573(97)00089-6

\bibitem{Jaffe:1996zw}
R.~L.~Jaffe,
[arXiv:hep-ph/9602236 [hep-ph]].

\bibitem{Burkardt:2002uc}
M.~Burkardt, X.~d.~Ji and F.~Yuan,
Phys. Lett. B \textbf{545}, 345-351 (2002)
doi:10.1016/S0370-2693(02)02596-0

\bibitem{Pasquini:2014ppa}
B.~Pasquini and P.~Schweitzer,
Phys. Rev. D \textbf{90}, no.1, 014050 (2014)
doi:10.1103/PhysRevD.90.014050

\bibitem{Metz:2016swz}
A.~Metz and A.~Vossen,
Prog. Part. Nucl. Phys. \textbf{91}, 136-202 (2016)
doi:10.1016/j.ppnp.2016.08.003

\bibitem{Pauli:1985ps}
H.~C.~Pauli and S.~J.~Brodsky,
Phys. Rev. D \textbf{32}, 2001 (1985)

\bibitem{Vary:2009gt}
J.~P.~Vary, H.~Honkanen, J.~Li, P.~Maris, S.~J.~Brodsky, A.~Harindranath, G.~F.~de Teramond, P.~Sternberg, E.~G.~Ng and C.~Yang,
Phys. Rev. C \textbf{81}, 035205 (2010)
doi:10.1103/PhysRevC.81.035205

\bibitem{Ji:2021znw}
X.~Ji and Y.~Liu,
Phys. Rev. D \textbf{105}, no.7, 076014 (2022)
doi:10.1103/PhysRevD.105.076014

\bibitem{Mezrag:2016hnp}
C.~Mezrag, H.~Moutarde and J.~Rodr\'iguez-Quintero,
Few Body Syst. \textbf{57}, no.9, 729-772 (2016)
doi:10.1007/s00601-016-1119-8

\bibitem{Shi:2018zqd}
C.~Shi and I.~C.~Clo\"et,
Phys. Rev. Lett. \textbf{122}, no.8, 082301 (2019)
doi:10.1103/PhysRevLett.122.082301

\bibitem{Eichmann:2021vnj}
G.~Eichmann, E.~Ferreira and A.~Stadler,
Phys. Rev. D \textbf{105}, no.3, 034009 (2022)
doi:10.1103/PhysRevD.105.034009

\bibitem{Bashir:2012fs}
A.~Bashir, L.~Chang, I.~C.~Clo\"et, B.~El-Bennich, Y.~X.~Liu, C.~D.~Roberts and P.~C.~Tandy,
Commun. Theor. Phys. \textbf{58}, 79-134 (2012)
doi:10.1088/0253-6102/58/1/16

\bibitem{Shi:2020pqe}
C.~Shi, K.~Bednar, I.~C.~Clo\"et and A.~Freese,
Phys. Rev. D \textbf{101}, no.7, 074014 (2020)
doi:10.1103/PhysRevD.101.074014

\bibitem{Kou:2023ady}
W.~Kou, C.~Shi, X.~Chen and W.~Jia,
Phys. Rev. D \textbf{108}, no.3, 036021 (2023)
doi:10.1103/PhysRevD.108.036021

\bibitem{Shi:2023oll}
C.~Shi, J.~Li, P.~L.~Yin and W.~Jia,
Phys. Rev. D \textbf{107}, no.7, 074009 (2023)
doi:10.1103/PhysRevD.107.074009

\bibitem{Almeida-Zamora:2023bqb}
B.~Almeida-Zamora, J.~J.~Cobos-Mart\'\i{}nez, A.~Bashir, K.~Raya, J.~Rodr\'\i{}guez-Quintero and J.~Segovia,
Phys. Rev. D \textbf{109}, no.1, 014016 (2024)
doi:10.1103/PhysRevD.109.014016

\bibitem{Rojas:2014aka}
E.~Rojas, B.~El-Bennich and J.~P.~B.~C.~de Melo,
Phys. Rev. D \textbf{90}, 074025 (2014)
doi:10.1103/PhysRevD.90.074025

\bibitem{Mojica:2017tvh}
F.~F.~Mojica, C.~E.~Vera, E.~Rojas and B.~El-Bennich,
Phys. Rev. D \textbf{96}, no.1, 014012 (2017)
doi:10.1103/PhysRevD.96.014012

\bibitem{Serna:2020txe}
F.~E.~Serna, R.~C.~da Silveira, J.~J.~Cobos-Mart\'\i{}nez, B.~El-Bennich and E.~Rojas,
Eur. Phys. J. C \textbf{80}, no.10, 955 (2020)
doi:10.1140/epjc/s10052-020-08517-3

\bibitem{Serna:2021xnr}
F.~E.~Serna and B.~El-Bennich,
PoS \textbf{CHARM2020}, 047 (2021)
doi:10.22323/1.385.0047

\bibitem{Serna:2022yfp}
F.~E.~Serna, R.~C.~da Silveira and B.~El-Bennich,
Phys. Rev. D \textbf{106}, no.9, L091504 (2022)
doi:10.1103/PhysRevD.106.L091504

\bibitem{daSilveira:2022pte}
R.~C.~da Silveira, F.~E.~Serna and B.~El-Bennich,
Phys. Rev. D \textbf{107}, no.3, 034021 (2023)
doi:10.1103/PhysRevD.107.034021

\bibitem{Llewellyn-Smith:1969bcu}
C.~H.~Llewellyn-Smith,
Annals Phys. \textbf{53}, 521-558 (1969)
doi:10.1016/0003-4916(69)90035-9

\bibitem{Serna:2018dwk}
F.~E.~Serna, C.~Chen and B.~El-Bennich,
Phys. Rev. D \textbf{99}, no.9, 094027 (2019)
doi:10.1103/PhysRevD.99.094027

\bibitem{Bhagwat:2002tx}
M.~Bhagwat, M.~A.~Pichowsky and P.~C.~Tandy,
Phys. Rev. D \textbf{67}, 054019 (2003)
doi:10.1103/PhysRevD.67.054019

\bibitem{El-Bennich:2016qmb}
B.~El-Bennich, G.~Krein, E.~Rojas and F.~E.~Serna,
Few Body Syst. \textbf{57}, no.10, 955-963 (2016)
doi:10.1007/s00601-016-1133-x

\bibitem{Mezrag:2015mka}
C.~Mezrag,
2015PA112144.

\bibitem{Ji:2003fw}
X.~d.~Ji, J.~P.~Ma and F.~Yuan,
Phys. Rev. Lett. \textbf{90}, 241601 (2003)
doi:10.1103/PhysRevLett.90.241601

\bibitem{Sutton:1991ay}
P.~J.~Sutton, A.~D.~Martin, R.~G.~Roberts and W.~J.~Stirling,
Phys. Rev. D \textbf{45}, 2349-2359 (1992)
doi:10.1103/PhysRevD.45.2349

\bibitem{Gluck:1999xe}
M.~Gluck, E.~Reya and I.~Schienbein,
Eur. Phys. J. C \textbf{10}, 313-317 (1999)
doi:10.1007/s100529900124

\bibitem{Botje:2010ay}
M.~Botje,
Comput. Phys. Commun. \textbf{182}, 490-532 (2011)
doi:10.1016/j.cpc.2010.10.020

\bibitem{dePaula:2022pcb}
W.~de Paula, E.~Ydrefors, J.~H.~Nogueira Alvarenga, T.~Frederico and G.~Salm\`e,
Phys. Rev. D \textbf{105}, no.7, L071505 (2022)
doi:10.1103/PhysRevD.105.L071505

\bibitem{Conway:1989fs}
J.~S.~Conway, C.~E.~Adolphsen, J.~P.~Alexander, K.~J.~Anderson, J.~G.~Heinrich, J.~E.~Pilcher, A.~Possoz, E.~I.~Rosenberg, C.~Biino and J.~F.~Greenhalgh, \textit{et al.}
Phys. Rev. D \textbf{39} (1989), 92-122
doi:10.1103/PhysRevD.39.92

\end{thebibliography}
\end{document}